\documentclass[a4paper, oneside, twocolumn, notitlepage, 10pt]{extarticle_ecoc}
\usepackage{ecoc}

\usepackage{bm}
\usepackage[nohyperlinks,nolist]{acronym}
\usepackage{pgfplots}
\usepackage{tikz}
\usepackage{pgffor}
\usepackage{array}
\usepackage{amsmath}

\usepgfplotslibrary{fillbetween}
\usetikzlibrary{positioning, chains, arrows.meta, shapes.geometric, bending, patterns.meta}

\definecolor{kit-green}{RGB}{0, 150, 130}
\colorlet{kit-green100}{kit-green}
\colorlet{kit-green90}{kit-green!90!white}
\colorlet{kit-green80}{kit-green!80!white}
\colorlet{kit-green70}{kit-green!70!white}
\colorlet{kit-green60}{kit-green!60!white}
\colorlet{kit-green50}{kit-green!50!white}
\colorlet{kit-green40}{kit-green!40!white}
\colorlet{kit-green30}{kit-green!30!white}
\colorlet{kit-green25}{kit-green!25!white}
\colorlet{kit-green20}{kit-green!20!white}
\colorlet{kit-green15}{kit-green!15!white}
\colorlet{kit-green10}{kit-green!10!white}
\colorlet{kit-green5}{kit-green!5!white}

\definecolor{kit-blue}{RGB}{70, 100, 170}
\colorlet{kit-blue100}{kit-blue}
\colorlet{kit-blue90}{kit-blue!90!white}
\colorlet{kit-blue80}{kit-blue!80!white}
\colorlet{kit-blue70}{kit-blue!70!white}
\colorlet{kit-blue60}{kit-blue!60!white}
\colorlet{kit-blue50}{kit-blue!50!white}
\colorlet{kit-blue40}{kit-blue!40!white}
\colorlet{kit-blue30}{kit-blue!30!white}
\colorlet{kit-blue25}{kit-blue!25!white}
\colorlet{kit-blue20}{kit-blue!20!white}
\colorlet{kit-blue15}{kit-blue!15!white}
\colorlet{kit-blue10}{kit-blue!10!white}
\colorlet{kit-blue5}{kit-blue!5!white}

\definecolor{kit-royalblue}{RGB}{0, 45, 76}
\colorlet{kit-royalblue100}{kit-royalblue}
\colorlet{kit-royalblue90}{kit-royalblue!90!white}
\colorlet{kit-royalblue80}{kit-royalblue!80!white}
\colorlet{kit-royalblue70}{kit-royalblue!70!white}
\colorlet{kit-royalblue60}{kit-royalblue!60!white}
\colorlet{kit-royalblue50}{kit-royalblue!50!white}
\colorlet{kit-royalblue40}{kit-royalblue!40!white}
\colorlet{kit-royalblue30}{kit-royalblue!30!white}
\colorlet{kit-royalblue25}{kit-royalblue!25!white}
\colorlet{kit-royalblue20}{kit-royalblue!20!white}
\colorlet{kit-royalblue15}{kit-royalblue!15!white}
\colorlet{kit-royalblue10}{kit-royalblue!10!white}
\colorlet{kit-royalblue5}{kit-royalblue!5!white}

\definecolor{kit-iceblue100}{RGB}{30, 53, 69}
\definecolor{kit-iceblue70}{RGB}{68, 94, 111}
\definecolor{kit-iceblue50}{RGB}{168, 185, 196}
\definecolor{kit-iceblue30}{RGB}{218, 225, 230}

\definecolor{kit-red}{RGB}{162, 34, 35}
\colorlet{kit-red100}{kit-red}
\colorlet{kit-red90}{kit-red!90!white}
\colorlet{kit-red80}{kit-red!80!white}
\colorlet{kit-red70}{kit-red!70!white}
\colorlet{kit-red60}{kit-red!60!white}
\colorlet{kit-red50}{kit-red!50!white}
\colorlet{kit-red40}{kit-red!40!white}
\colorlet{kit-red30}{kit-red!30!white}
\colorlet{kit-red25}{kit-red!25!white}
\colorlet{kit-red20}{kit-red!20!white}
\colorlet{kit-red15}{kit-red!15!white}
\colorlet{kit-red10}{kit-red!10!white}
\colorlet{kit-red5}{kit-red!5!white}

\definecolor{kit-yellow}{RGB}{252, 229, 0}
\colorlet{kit-yellow100}{kit-yellow}
\colorlet{kit-yellow90}{kit-yellow!90!white}
\colorlet{kit-yellow80}{kit-yellow!80!white}
\colorlet{kit-yellow70}{kit-yellow!70!white}
\colorlet{kit-yellow60}{kit-yellow!60!white}
\colorlet{kit-yellow50}{kit-yellow!50!white}
\colorlet{kit-yellow40}{kit-yellow!40!white}
\colorlet{kit-yellow30}{kit-yellow!30!white}
\colorlet{kit-yellow25}{kit-yellow!25!white}
\colorlet{kit-yellow20}{kit-yellow!20!white}
\colorlet{kit-yellow15}{kit-yellow!15!white}
\colorlet{kit-yellow10}{kit-yellow!10!white}
\colorlet{kit-yellow5}{kit-yellow!5!white}

\definecolor{kit-orange}{RGB}{223, 155, 27}
\colorlet{kit-orange100}{kit-orange}
\colorlet{kit-orange90}{kit-orange!90!white}
\colorlet{kit-orange80}{kit-orange!80!white}
\colorlet{kit-orange70}{kit-orange!70!white}
\colorlet{kit-orange60}{kit-orange!60!white}
\colorlet{kit-orange50}{kit-orange!50!white}
\colorlet{kit-orange40}{kit-orange!40!white}
\colorlet{kit-orange30}{kit-orange!30!white}
\colorlet{kit-orange25}{kit-orange!25!white}
\colorlet{kit-orange20}{kit-orange!20!white}
\colorlet{kit-orange15}{kit-orange!15!white}
\colorlet{kit-orange10}{kit-orange!10!white}
\colorlet{kit-orange5}{kit-orange!5!white}

\definecolor{kit-lightgreen}{RGB}{140, 182, 60}
\colorlet{kit-lightgreen100}{kit-lightgreen}
\colorlet{kit-lightgreen90}{kit-lightgreen!90!white}
\colorlet{kit-lightgreen80}{kit-lightgreen!80!white}
\colorlet{kit-lightgreen70}{kit-lightgreen!70!white}
\colorlet{kit-lightgreen60}{kit-lightgreen!60!white}
\colorlet{kit-lightgreen50}{kit-lightgreen!50!white}
\colorlet{kit-lightgreen40}{kit-lightgreen!40!white}
\colorlet{kit-lightgreen30}{kit-lightgreen!30!white}
\colorlet{kit-lightgreen25}{kit-lightgreen!25!white}
\colorlet{kit-lightgreen20}{kit-lightgreen!20!white}
\colorlet{kit-lightgreen15}{kit-lightgreen!15!white}
\colorlet{kit-lightgreen10}{kit-lightgreen!10!white}
\colorlet{kit-lightgreen5}{kit-lightgreen!5!white}

\definecolor{kit-purple}{RGB}{163, 16, 124}
\colorlet{kit-purple100}{kit-purple}
\colorlet{kit-purple90}{kit-purple!90!white}
\colorlet{kit-purple80}{kit-purple!80!white}
\colorlet{kit-purple70}{kit-purple!70!white}
\colorlet{kit-purple60}{kit-purple!60!white}
\colorlet{kit-purple50}{kit-purple!50!white}
\colorlet{kit-purple40}{kit-purple!40!white}
\colorlet{kit-purple30}{kit-purple!30!white}
\colorlet{kit-purple25}{kit-purple!25!white}
\colorlet{kit-purple20}{kit-purple!20!white}
\colorlet{kit-purple15}{kit-purple!15!white}
\colorlet{kit-purple10}{kit-purple!10!white}
\colorlet{kit-purple5}{kit-purple!5!white}

\definecolor{kit-brown}{RGB}{167, 130, 46}
\colorlet{kit-brown100}{kit-brown}
\colorlet{kit-brown90}{kit-brown!90!white}
\colorlet{kit-brown80}{kit-brown!80!white}
\colorlet{kit-brown70}{kit-brown!70!white}
\colorlet{kit-brown60}{kit-brown!60!white}
\colorlet{kit-brown50}{kit-brown!50!white}
\colorlet{kit-brown40}{kit-brown!40!white}
\colorlet{kit-brown30}{kit-brown!30!white}
\colorlet{kit-brown25}{kit-brown!25!white}
\colorlet{kit-brown20}{kit-brown!20!white}
\colorlet{kit-brown15}{kit-brown!15!white}
\colorlet{kit-brown10}{kit-brown!10!white}
\colorlet{kit-brown5}{kit-brown!5!white}

\definecolor{kit-cyan}{RGB}{35, 161, 224}
\colorlet{kit-cyan100}{kit-cyan}
\colorlet{kit-cyan90}{kit-cyan!90!white}
\colorlet{kit-cyan80}{kit-cyan!80!white}
\colorlet{kit-cyan70}{kit-cyan!70!white}
\colorlet{kit-cyan60}{kit-cyan!60!white}
\colorlet{kit-cyan50}{kit-cyan!50!white}
\colorlet{kit-cyan40}{kit-cyan!40!white}
\colorlet{kit-cyan30}{kit-cyan!30!white}
\colorlet{kit-cyan25}{kit-cyan!25!white}
\colorlet{kit-cyan20}{kit-cyan!20!white}
\colorlet{kit-cyan15}{kit-cyan!15!white}
\colorlet{kit-cyan10}{kit-cyan!10!white}
\colorlet{kit-cyan5}{kit-cyan!5!white}

\definecolor{kit-gray}{RGB}{0, 0, 0}
\colorlet{kit-gray100}{kit-gray}
\colorlet{kit-gray90}{kit-gray!90!white}
\colorlet{kit-gray80}{kit-gray!80!white}
\colorlet{kit-gray70}{kit-gray!70!white}
\colorlet{kit-gray60}{kit-gray!60!white}
\colorlet{kit-gray50}{kit-gray!50!white}
\colorlet{kit-gray40}{kit-gray!40!white}
\colorlet{kit-gray30}{kit-gray!30!white}
\colorlet{kit-gray25}{kit-gray!25!white}
\colorlet{kit-gray20}{kit-gray!20!white}
\colorlet{kit-gray15}{kit-gray!15!white}
\colorlet{kit-gray10}{kit-gray!10!white}
\colorlet{kit-gray5}{kit-gray!5!white}

\acrodef{GMI}[GMI]{generalized mutual information}
\acrodef{CNN}[CNN]{convolutional neural network}
\acrodef{PAM}[PAM]{pulse-amplitude modulation}
\acrodef{RRC}[RRC]{root-raised cosine}
\acrodef{rvmb}[rvmb]{real-valued multiplications per bit}
\acrodef{NGMI}[NGMI]{normalized generalized mutual information}
\acrodef{DM}[DM]{distribution matcher}
\acrodef{IMDD}[IM/DD]{intensity modulation with direct detection}
\acrodef{CD}[CD]{chromatic dispersion}
\acrodef{RL}[RL]{reinforcement learning}
\acrodef{DSP}[DSP]{digital signal processing}
\acrodef{ACF}[ACF]{autocorrelation function}
\acrodef{PSD}[PSD]{power spectral density}
\acrodef{DD}[DD]{direct detection}
\acrodef{CS}[CS]{constellation shaping}
\acrodef{LLR}[LLR]{log-likelihood ratio}
\acrodef{ASE}[ASE]{amplified spontaneous emission}

\begin{document}
\selectlanguage{english}    %

\title{Joint Probabilistic and Geometric Constellation Shaping for Complexity-Constrained Direct Detection Optical Systems}%

\author{
    Rodrigo Fischer\textsuperscript{(1)}, Shrinivas Chimmalgi\textsuperscript{(1)},
    Andrej Rode\textsuperscript{(1)}, Laurent Schmalen\textsuperscript{(1)}
}

\maketitle                  %

\begin{strip}
    \begin{author_descr}

        \textsuperscript{(1)} Communications Engineering Lab (CEL), Karlsruhe Institute of Technology (KIT)
        
        * Corresponding author: \textcolor{blue}{\uline{rodrigo.fischer@kit.edu}}
        
    \end{author_descr}
\end{strip}

\renewcommand\footnotemark{}
\renewcommand\footnoterule{}

\begin{strip}
    \begin{ecoc_abstract}
        We evaluate joint probabilistic and geometric constellation shaping via reinforcement learning for complexity-constrained joint equalization and demodulation of direct detection optical signals. We demonstrate the proposed technique in a simulated 56~GBd, 2.2~km C-band direct-detection system, demonstrating its effectiveness for complexity-constrained receivers. ©2026 The Author(s) 
    \end{ecoc_abstract}
\end{strip}

\section{Introduction}

While \ac{DD} remains the preferred technology for short-reach optical links due to its advantages in cost and power consumption~\cite{imdd}, it can introduce significant impairments to the signal. Specifically, \ac{DD} combined with the effect of \ac{CD} accumulated over the fiber results in an equivalent non-linear channel with memory. To mitigate this, various strategies have been explored; notably, machine learning-based non-linear \ac{DSP} has been evaluated for low-complexity equalization in \ac{DD} links~\cite{kan, murphy2023, Lauinger:24}.

At the same time, \ac{CS} is required to close the gap to channel capacity~\cite{pas} and has already been assessed for \ac{DD} systems~\cite{chagnon, chimmalgi}. However, constraining the computational complexity of the \ac{DSP} at the receiver impacts the achievable information rate in the form of an equivalent degraded channel, and the analysis of the effect of such constraints on the shaped constellations remains an open challenge.  

In this work, we use soft thresholding algorithms~\cite{doi:10.1137/080716542} in order to constrain the computational complexity of \ac{CNN}-based joint equalizers and demodulators, combined with \ac{RL}~\cite{Williams1992} techniques in order to perform joint probabilistic and geometric constellation shaping via the \ac{GMI}~\cite{Kaplan1993228}, a metric that is able to take the degradation introduced by the constrained receiver into account during shaping. 

\section{System Model}

Given the set of symbol probabilities ${\mathcal{P}=\{p_1,\ldots,p_M\}}$, with $M=2^m$, the symbol indices $S_n$ are sampled with $P_{S_n}(i)=p_i$. Given an alphabet $\mathcal{C}=\{c_1, \ldots, c_{M}\}\subset\mathbb{R}$, the transmitted symbols with normalized energy are given by $X_n=c_{S_n} / \sqrt{\mathbb{E}[|c_{S_n}|^2]}$. Following Fig.~\ref{fig:diagram}, $X_n$ are then modulated using a \ac{RRC} filter with roll-off factor $\beta$. The signal is then fed to the fiber, where \ac{CD} accumulates. Noise stemming from \ac{ASE} is modeled as complex AWGN with \ac{PSD} $\nu_\text{opt}$, and electrical noise after \ac{DD} is modeled as AWGN $N_\text{el}(t)$ with \ac{PSD} $\nu_\text{el}$. \Ac{DD} is modeled by an absolute squared operation $|\cdot|^2$, and in order to limit the noise bandwidth, we introduce brickwall filters $H_\text{PD}(f), H_\text{RX}(f)$ with bandwidths $B_\text{PD}=(1+\beta)/T$ and $B_\text{RX}=2(1+\beta)/T$. The received signal $Y(t)$ is sampled at a fixed rate of $N_\text{sps}=2$.

We pass the received samples through a \ac{CNN} that performs joint equalization and demodulation, referred here as the \textit{detector}. At each symbol time step, the \ac{CNN} outputs for each bit $1\leq i \leq m$ the \ac{LLR} estimates $L_i(\bm{Y}_n; \bm{\theta})$, where $\bm{\theta}$ are the tunable parameters of the detector and ${\bm{Y}_n=(Y_{2n-L_\text{CNN}/2},\ldots,Y_{2n},\ldots,Y_{2n+L_\text{CNN}/2})}$ are the $Y_k$ samples that are used for the $n$th symbol \ac{LLR} estimates.

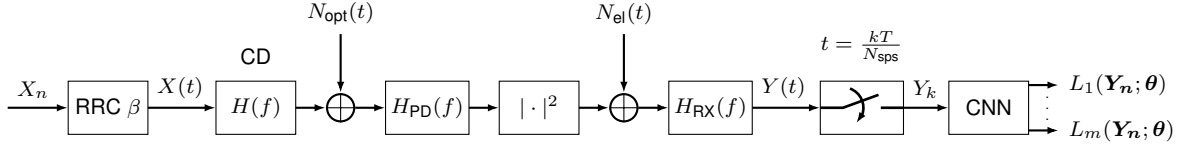
\begin{figure*}
    \centering
    \begin{tikzpicture}[
    node distance = 0.9cm,
    font=\footnotesize,
    on grid = false,
    block/.style = {draw, minimum height=2em, minimum width=3em, outer sep=0pt, inner sep=2pt},
    adder/.style = {draw, circle, minimum size=0.36cm, inner sep=0pt, thick},
    every join/.style = {->, thick},
    start chain = going right,
    >={Latex[length=1.5mm, width=1mm]}
]

    \node [on chain, inner sep=0] (start) {};
    \node [block, on chain, join, right=0.8cm of start] (dac) {RRC $\beta$};
    \node [block, on chain, join] (chan) {$H(f)$};
    
    \node [adder, on chain, join, right=0.4cm of chan] (p1) {};
    \draw (p1.north) -- (p1.south) (p1.west) -- (p1.east);
    
    \node [block, on chain, join, right=0.4cm of p1] (fd) {$H_\text{PD}(f)$};
    \node [block, on chain, join, right=0.4cm of fd] (abs) {$| \cdot |^2$};
    
    \node [adder, on chain, join, right=0.4cm of abs] (p2) {};
    \draw (p2.north) -- (p2.south) (p2.west) -- (p2.east);
    
    \node [block, on chain, join, right=0.4cm of p2] (frx) {$H_\text{RX}(f)$};
    
    \node [block, on chain, join, minimum width=1.1cm] (samp) {};
    \node [block, on chain, join, right=0.6cm of samp] (demod) {CNN};

    \foreach \p/\txt in {p1/N_\text{opt}(t), p2/N_\text{el}(t)} {
        \node [above=0.78cm of \p] (n-\p) {$\txt$};
        \draw [->, thick] (n-\p) -- (\p);
    }

    \begin{scope}[line width=1]
        \draw (samp.west) -- ++(0.3, 0) coordinate (sw_pivot) -- ++(20:0.5);
        \draw (samp.west) ++(0.8, 0) -- (samp.east);
        \draw [->] (sw_pivot) ++(70:0.27) arc (70:-50:0.27);
    \end{scope}

    \draw [->, thick] (demod.east) ++(0, 0.3cm) -- ++(0.4cm, 0cm) node [anchor=west] {$L_1(\bm{Y_n}; \bm{\theta})$};
    \draw [->, thick] (demod.east) ++(0, -0.3cm) -- ++(0.4cm, 0cm) node [anchor=west] {$L_m(\bm{Y_n}; \bm{\theta})$};
    \path (demod.east) +(0.1cm, 0.1cm) node[anchor=west, font=\tiny] {$\vdots$};

    \node [anchor=south west] at (start.east) {$X_n$};
    \node [above=0.1cm of samp] {$t=\frac{kT}{N_\text{sps}}$};
    \node [above=0.1cm of chan] {CD};
    \node [anchor=south west] at (dac.east) {$X(t)$};
    \node [anchor=south west] at (frx.east) {$Y(t)$};
    \node [anchor=south west] at (samp.east) {$Y_k$};

\end{tikzpicture}%
    \caption{System model with chromatic dispersion, direct detection and \ac{CNN}-based joint equalization and demapping.}
    \label{fig:diagram}
\end{figure*}

\section{Optimization Metric}

We optimize the detector, the symbol probabilities and the alphabet using the \ac{GMI}~\cite{Kaplan1993228} of the bits associated with the $n$th symbol $\bm{B}_n$ and the samples used for detection $\bm{Y}_n$
\begin{equation*}
    I_{\text{GMI}}(\bm{B_n}; \bm{Y_n}) = \sup_{s>0} \underset{\bm{B}_n\!,\!\bm{Y}_n}{\mathbb{E}}\left[\!\log \frac{Q(\bm{B}_n, \bm{Y}_n; \bm{\theta})^s}{\underset{\bm{B}_n'}{\mathbb{E}}\left[ Q(\bm{B}_n', \bm{Y} ; \bm{\theta})^s \right]} \! \right]
\end{equation*}
for a detector $Q(\bm{b}, \bm{y} ; \bm{\theta})$ parameterized by $\bm{\theta}$. For bit-wise decoding, $Q(\bm{b}, \bm{y} ; \bm{\theta}) = \prod_{i=1}^{m}q_i(b_i, \bm{y} ; \bm{\theta})$, and we can rearrange the terms to obtain $I_{\text{GMI}}(\bm{B}_n, \bm{Y}_n) = \sup_{s>0} \mathbb{E}\left[ \lambda(\bm{B}_n, \bm{Y}_n ; \bm{\theta})\right]$, where $\lambda(\bm{b}, \bm{y}; \bm{\theta})$ is given by
\begin{equation*}
    - \log \! \sum_{\bm{b}' \in \mathcal{B}_m} \!\! \exp \!\! \left(\ln P(\bm{b'}) + s \sum_{i=1}^m(b_i - b_i')L_i(\bm{y};\bm{\theta})\right),
\end{equation*}
where $L_i(\bm{y}; \bm{\theta})=\ln(q_i(0, \bm{y};\bm{\theta}) / q_i(1, \bm{y};\bm{\theta}))$ is the \ac{LLR} estimate for the $i$th bit. \acp{CNN} can incorporate the scaling $s$ on the last layer of $L_i(\bm{y}; \bm{\theta})$, so we drop that from the optimization objective without loss of generality. 

\begin{figure}[b]
    \centering
    \begin{tikzpicture}

\begin{axis}[
width=3.3cm,
height=4cm,
at={(0,0)},
scale only axis,
ytick distance=1,
minor y tick num=4,
ylabel=GMI,
ylabel style={at={(0.30, 0.5)}, anchor=near ticklabel},
ymin=0.6,
ymax=3.1,
xmax=1e5,
xlabel=Complexity (rvmb),
xmode=log,
xmin=2e-1,
xminorticks=true,
minor x tick num=8,
xtick={0.1, 1, 10, 1e2, 1e3, 1e4, 1e5},
xticklabels={,$10^0$,,$10^2$,,$10^4$,},
xlabel style={at={(0.50, 0.05)}, anchor=near ticklabel},
grid=major,
minor grid style={thin, gray!20},
no marks,
legend style={
    at={(1.015, 1.01)},
    anchor=south,
    legend columns=2,
    align=left,
    inner sep=1pt,
    draw=none,
    /tikz/every even column/.append style={column sep=0.65cm},
    font=\footnotesize,
},
]

\addlegendimage{black!80, line width=1, densely dashed}
\addlegendimage{black, line width=1}
\addlegendentry{$E_s/\nu_\text{opt}$~$=X$~dB\\\rlap{$E_s/\nu_\text{el}$}\phantom{$E_s/\nu_\text{opt}$}~$=\infty$}
\addlegendentry{$E_s/\nu_\text{opt}$~$=\infty$\\\rlap{$E_s/\nu_\text{el}$}\phantom{$E_s/\nu_\text{opt}$}~$=X$~dB}

\node[anchor=south west, fill=white, font=\footnotesize, inner sep=2pt] at (axis cs: 0.2, 2.8) {a) $X=15$~dB};

\addplot [draw=kit-purple, densely dashed, line width=1] table [
    x=x, y=mean,
] {figures/data/pareto_2_15_inf.dat};

\addplot [draw=kit-cyan, densely dashed, line width=1] table [
    x=x, y=mean,
] {figures/data/pareto_4_15_inf.dat};

\addplot [draw=kit-green, densely dashed, line width=1] table [
    x=x, y=mean,
] {figures/data/pareto_8_15_inf.dat};

\addplot [draw=kit-purple!70!black, line width=1] table [
    x=x, y=mean,
] {figures/data/pareto_2_inf_15.dat};

\addplot [draw=kit-cyan!70!black, line width=1] table [
    x=x, y=mean,
] {figures/data/pareto_4_inf_15.dat};

\addplot [draw=kit-green!70!black, line width=1] table [
    x=x, y=mean,
] {figures/data/pareto_8_inf_15.dat};

\node[draw, ellipse, minimum width=0.1cm, minimum height=0.3cm, inner sep=0] (E) at (axis cs: 3e3, 2.13) {}; 
\node[font=\scriptsize, inner sep=2pt, fill=white] (8pam) at (axis cs: 2e2, 2.4) {8PAM};
\draw (8pam.east) -- (E); 
\node[draw, ellipse, minimum width=0.1cm, minimum height=0.3cm, inner sep=0] (E) at (axis cs: 3e3, 1.86) {}; 
\node[font=\scriptsize, inner sep=2pt, fill=white] (4pam) at (axis cs: 2e2, 1.6) {4PAM};
\draw (4pam.east) -- (E); 
\node[draw, ellipse, minimum width=0.1cm, minimum height=0.3cm, inner sep=0] (E) at (axis cs: 3e3, 1) {}; 
\node[font=\scriptsize, inner sep=2pt, fill=white] (2pam) at (axis cs: 2e2, 1.2) {2PAM};
\draw (2pam.east) -- (E); 

\end{axis}

\begin{axis}[
width=3.3cm,
height=4cm,
at={(3.4cm,0)},
scale only axis,
ytick distance=1,
minor y tick num=4,
yticklabels={},
ymin=0.6,
ymax=3.1,
xmax=1e5,
xlabel=Complexity (rvmb),
xmode=log,
xmin=2e-1,
xminorticks=true,
minor x tick num=8,
xtick={0.1, 1, 10, 1e2, 1e3, 1e4, 1e5},
xticklabels={,$10^0$,,$10^2$,,$10^4$,},
xlabel style={at={(0.50, 0.05)}, anchor=near ticklabel},
grid=major,
minor grid style={thin, gray!20},
no marks,
legend style={
    at={(1,0.05)},
    anchor=south east
},
]

\node[anchor=south west, fill=white, font=\footnotesize, inner sep=2pt] at (axis cs: 0.2, 2.8) {b) $X=25$~dB};

\addplot [draw=kit-purple, densely dashed, line width=1] table [
    x=x, y=mean,
] {figures/data/pareto_2_25_inf.dat};

\addplot [draw=kit-cyan, densely dashed, line width=1] table [
    x=x, y=mean,
] {figures/data/pareto_4_25_inf.dat};

\addplot [draw=kit-green, densely dashed, line width=1] table [
    x=x, y=mean,
] {figures/data/pareto_8_25_inf.dat};

\addplot [draw=kit-purple!70!black, line width=1] table [
    x=x, y=mean,
] {figures/data/pareto_2_inf_25.dat};

\addplot [draw=kit-cyan!70!black, line width=1] table [
    x=x, y=mean,
] {figures/data/pareto_4_inf_25.dat};

\addplot [draw=kit-green!70!black, line width=1] table [
    x=x, y=mean,
] {figures/data/pareto_8_inf_25.dat};

\node[draw, ellipse, minimum width=0.1cm, minimum height=0.3cm, inner sep=0] (E) at (axis cs: 3e3, 2.90) {}; 
\node[font=\scriptsize, inner sep=2pt, fill=white] (8pam) at (axis cs: 1e4, 2.6) {8PAM};
\draw (8pam.north) -- (E); 
\node[draw, ellipse, minimum width=0.1cm, minimum height=0.3cm, inner sep=0] (E) at (axis cs: 3e3, 2) {}; 
\node[font=\scriptsize, inner sep=2pt, fill=white] (4pam) at (axis cs: 1e4, 1.8) {4PAM};
\draw (4pam.north) -- (E); 
\node[draw, ellipse, minimum width=0.1cm, minimum height=0.3cm, inner sep=0] (E) at (axis cs: 3e3, 1) {}; 
\node[font=\scriptsize, inner sep=2pt, fill=white] (2pam) at (axis cs: 1e4, 1.2) {2PAM};
\draw (2pam.south) -- (E); 

\end{axis}

\end{tikzpicture}%
    \caption{Pareto fronts of GMI$\times$rvmb for 2,4,8PAM at two different noise levels, namely 15~dB and 25~dB, applied either in the optical or electrical domain.}
    \label{fig:paretos_snr}
\end{figure}
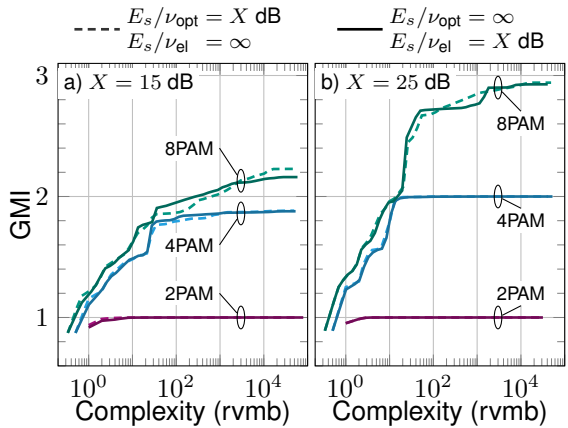

\section{Training the Detector}

We optimize $\bm{\theta}$ via proximal gradient descent~\cite{doi:10.1137/080716542}, where a $L_1$ regularization penalty is introduced in order to induce sparsity in $\bm{\theta}$. Setting fewer parameters to nonzero reduces the complexity of the detector in terms of real valued multiplications performed, and also in terms of the memory required to store the parameters. 

\section{Constellation Shaping}

We define probabilistic shaping when we optimize $\mathcal{P}$, geometric shaping when we optimize $\mathcal{C}$ and joint shaping when we optimize $\mathcal{C}$ and $\mathcal{P}$. Since $\mathcal{C}$ and $\mathcal{P}$ affect the distribution of $\bm{B}, \bm{Y}$, we cannot interchange expectation with differentiation~\cite{chagnon}. Instead, we have to use either a continuous relaxation of the sampling process of $S_n$ based on the Gumbel distribution~\cite{gumbel}, or use \ac{RL} approaches~\cite{Williams1992}. We chose the latter, and the first step is to optimize the logits $\bm{\ell}=(\ell_1, \ldots,\ell_M)$ instead, which are related to the probabilities by $p_i=\text{softmax}_i(\bm{\ell})$. The total gradients are given by $\nabla_{\ell_i}\mathbb{E}[\lambda] = \mathbb{E}[\nabla_{\ell_i}\lambda + \nabla_{\ell_i, \text{score}}]$, where the extra score term is given by~\cite{Williams1992}
\begin{equation*}
    \nabla_{l_i,\text{score}} = \lambda(\bm{b}_n, \bm{y}_n; \bm{\theta})\left(\#_i(\bm{y}_n) - p_i\frac{L_\text{CNN}+1}{N_\text{sps}}\right),
\end{equation*}
where we count how many times the symbol with index $i$ appeared in the symbol sequence that generated $\bm{y}_n$, given by $\#_i(\bm{y}_n)$, and we subtract by the expected occurrence of this symbol $p_i(L_\text{CNN}+1)/N_\text{sps}$. If a particular sequence has more occurrences of index $i$ than $p_i$ predicts on average, and also happens to have a better metric $\lambda(\bm{b}_n, \bm{y}_n;\bm{\theta})$, then the score term will increase the value of $l_i$.

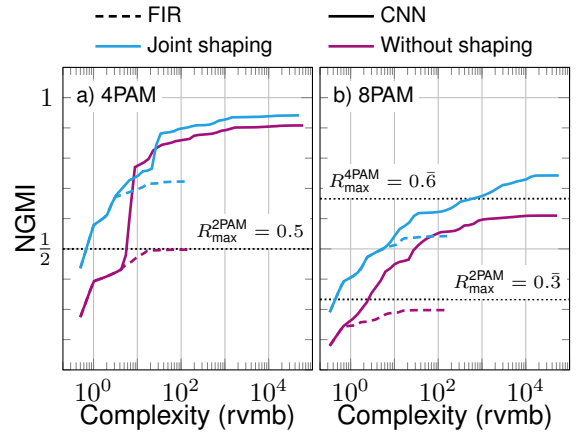
\begin{figure}[b]
    \centering
    \begin{tikzpicture}

\begin{axis}[
width=3.3cm,
height=4cm,
at={(0,0)},
scale only axis,
ytick distance=1,
minor y tick num=4,
ylabel=NGMI,
ylabel style={at={(0.30, 0.5)}, anchor=near ticklabel},
ymin=0.1,
ymax=1.1,
ytick={0, 0.5, 1, 1.5},
yticklabels={,$\frac{1}{2}$,$1$,},
xmax=1e5,
xlabel=Complexity (rvmb),
xmode=log,
xmin=2e-1,
xminorticks=true,
minor x tick num=8,
xtick={0.1, 1, 10, 1e2, 1e3, 1e4, 1e5},
xticklabels={,$10^0$,,$10^2$,,$10^4$,},
xlabel style={at={(0.50, 0.05)}, anchor=near ticklabel},
grid=major,
minor grid style={thin, gray!20},
no marks,
legend style={
    at={(1.015, 1.01)},
    anchor=south,
    legend columns=2,
    legend cell align={left},
    inner sep=1pt,
    draw=none,
    /tikz/every even column/.append style={column sep=0.65cm},
    font=\footnotesize
},
]

\addlegendimage{black, line width=1, densely dashed}
\addlegendimage{black, line width=1}
\addlegendimage{kit-cyan, line width=1}
\addlegendimage{kit-purple, line width=1}
\addlegendentry{FIR}
\addlegendentry{CNN}
\addlegendentry{Joint shaping}
\addlegendentry{Without shaping}

\node[anchor=south west, fill=white, font=\footnotesize, inner sep=2pt] at (axis cs: 0.3, 0.95) {a) 4PAM};

\draw [kit-gray, line width=0.7, font=\scriptsize, densely dotted, anchor=south east] (axis cs:2e-1, 0.5) -- (axis cs:1e5, 0.5)
    node[pos=0.75, above, fill=white, inner sep=2pt] {$R_\text{max}^\text{2PAM}=0.5$};

\addplot [draw=kit-purple, densely dashed, line width=1] table [
    x=x, y=mean,
] {figures/data/pareto_false_false_4_fir.dat};

\addplot [draw=kit-purple, line width=1] table [
    x=x, y=mean,
] {figures/data/pareto_false_false_4.dat};

\addplot [draw=kit-cyan, densely dashed, line width=1] table [
    x=x, y=mean,
] {figures/data/pareto_true_true_4_fir.dat};

\addplot [draw=kit-cyan, line width=1] table [
    x=x, y=mean,
] {figures/data/pareto_true_true_4.dat};

\end{axis}

\begin{axis}[
width=3.3cm,
height=4cm,
at={(3.4cm,0)},
scale only axis,
ytick distance=1,
minor y tick num=4,
ytick={0, 0.5, 1, 1.5},
yticklabels={},
ymin=0.1,
ymax=1.1,
xmax=1e5,
xlabel=Complexity (rvmb),
xmode=log,
xmin=2e-1,
xminorticks=true,
minor x tick num=8,
xtick={0.1, 1, 10, 1e2, 1e3, 1e4, 1e5},
xticklabels={,$10^0$,,$10^2$,,$10^4$,},
xlabel style={at={(0.50, 0.05)}, anchor=near ticklabel},
grid=major,
minor grid style={thin, gray!20},
no marks,
legend style={
    at={(1,0.05)},
    anchor=south east
},
]

\node[anchor=south west, fill=white, font=\footnotesize, inner sep=2pt] at (axis cs: 0.3, 0.95) {b) 8PAM};

\draw [kit-gray, line width=0.7, font=\scriptsize, densely dotted, anchor=south east] (axis cs:2e-1, 0.333) -- (axis cs:1e5, 0.333)
    node[pos=0.75, above, fill=white, inner sep=2pt] {$R_\text{max}^\text{2PAM}=0.\bar{3}$};
\draw [kit-gray, line width=0.7, font=\scriptsize, densely dotted, anchor=south west] (axis cs:2e-1, 0.666) -- (axis cs:1e5, 0.666)
    node[pos=0.25, above, fill=white, inner sep=2pt] {$R_\text{max}^\text{4PAM}=0.\bar{6}$};

\addplot [draw=kit-purple, densely dashed, line width=1] table [
    x=x, y=mean,
] {figures/data/pareto_false_false_8_fir.dat};

\addplot [draw=kit-purple, line width=1] table [
    x=x, y=mean,
] {figures/data/pareto_false_false_8.dat};

\addplot [draw=kit-cyan, densely dashed, line width=1] table [
    x=x, y=mean,
] {figures/data/pareto_true_true_8_fir.dat};

\addplot [draw=kit-cyan, line width=1] table [
    x=x, y=mean,
] {figures/data/pareto_true_true_8.dat};

\end{axis}

\end{tikzpicture}%
    \caption{Pareto fronts of NGMI$\times$rvmb for 4,8PAM with joint shaping and without shaping for FIR filters and for $n$-layer \acp{CNN}.}
    \label{fig:paretos_ngmi}
\end{figure}

\section{Simulation Results}

We simulated a 2.2~km long fiber with a dispersion coefficient of $\text{16.3}\,\text{ps}\cdot\text{nm}^{-1}\cdot\text{km}^{-1}$ and a carrier wavelength of 1540~nm. We considered 2,4,8PAM signals modulated at 56~GBd with symbols ${\mathcal{C}\subset\mathbb{R}}$; bipolar signals can be generated with a single Mach-Zehnder modulator for the in-phase component.

We used a batch size of 64 sequences and a batch length of 512 symbols. We perform filtering in the frequency domain, meaning all convolutions are cyclic. In order to simulate the full bandwidth of the signals, we used a simulation oversampling factor of $N_\text{os,sim}=4$. We limited the number of gradient steps to 8000, with a grace period of 1000 steps where $\mathcal{C}, \mathcal{P}$ are not updated, only the detector. We found this to be vital for convergence, since we want to avoid $\mathcal{C}, \mathcal{P}$ to adapt to a poorly fitted detector.

We use the \ac{GMI} as a performance metric and the number of \ac{rvmb} as a computational complexity metric. In order to explore a large range of complexities, we trained an ensemble of \ac{CNN} architectures with 1 to 4 layers with varying kernel sizes for each layer and stored the achieved \ac{GMI}, generating a 2D set suitable for multi-objective optimization, from which we extract the Pareto fronts.

In Fig.~\ref{fig:paretos_snr} we optimize $\mathcal{C}, \mathcal{P}, L_i(\cdot,\bm{\theta})$ for the case where either the optical or electrical noise dominates. We find that under our simulation settings, where the detector complexity is constrained, the performance is dictated only by the noise level and doesn't change if $\nu_\text{el}=0$ or $\nu_\text{opt}=0$. For the remaining of the simulations, we consider $E_s/\nu_\text{opt}=15$~dB and $\nu_\text{el}=0$.

Next, in Fig.~\ref{fig:paretos_ngmi} we compare the \ac{NGMI}~\cite{ngmi_schmal, ngmi_winzer}, for 4,8PAM considering joint shaping and also without shaping. For 8PAM, we observe a \ac{NGMI} gap of approximately 10\% throughout the entire complexity range of the detector, whereas for 4PAM, we observe a gap of approximately 5\% starting at 100~\ac{rvmb}. As a reference, we also plot the Pareto fronts for FIR filters, where we see that for high-rate communications, linear processing is not sufficient for good performance. For 8PAM, we only achieve rates that are not achievable by 4PAM if the \ac{NGMI} exceed the $R_\text{max}^\text{4PAM}$ threshold, which occurs at around $10^3$~rvmb, only reached by multi-layer \acp{CNN}. For 4PAM, we reach an \ac{NGMI} of 0.9 at around 100~rvmb joint using shaping.

\begin{figure}
    \centering
    \begin{tikzpicture}

\begin{axis}[
width=3.3cm,
height=4cm,
at={(0,0)},
scale only axis,
ytick distance=1,
minor y tick num=4,
ylabel=NGMI,
ylabel style={at={(0.30, 0.5)}, anchor=near ticklabel},
ymin=0.1,
ymax=1.1,
ytick={0, 0.5, 1, 1.5},
yticklabels={,$\frac{1}{2}$,$1$,},
xmax=1e5,
xlabel=Complexity (rvmb),
xmode=log,
xmin=2e-1,
xminorticks=true,
minor x tick num=8,
xtick={0.1, 1, 10, 1e2, 1e3, 1e4, 1e5},
xticklabels={,$10^0$,,$10^2$,,$10^4$,},
xlabel style={at={(0.50, 0.05)}, anchor=near ticklabel},
grid=major,
minor grid style={thin, gray!20},
no marks,
legend style={
    at={(1.015, 1.01)},
    anchor=south,
    legend columns=2,
    legend cell align={left},
    inner sep=1pt,
    draw=none,
    /tikz/every even column/.append style={column sep=0.15cm},
    font=\footnotesize
},
]

\addlegendimage{black, line width=1, densely dashed}
\addlegendimage{black, line width=1}
\addlegendimage{kit-cyan, line width=1}
\addlegendimage{kit-purple, line width=1}
\addlegendentry{FIR}
\addlegendentry{CNN}
\addlegendentry{Geometric shaping}
\addlegendentry{Probabilistic shaping}

\node[anchor=south west, fill=white, inner sep=2pt] at (axis cs: 0.3, 0.95) {\footnotesize a) 4PAM};

\node[font=\scriptsize, fill=white, inner sep=2pt, align=center] (true) at (axis cs: 2, 0.85) {Joint\\shaping};

\draw [kit-gray, line width=0.7, font=\scriptsize, densely dotted, anchor=south east] (axis cs:2e-1, 0.5) -- (axis cs:1e5, 0.5)
    node[pos=0.75, above, fill=white, inner sep=2pt] {$R_\text{max}^\text{2PAM}=0.5$};

\addplot [draw=black!30, line width=1] table [
    x=x, y=mean,
] {figures/data/pareto_true_true_4.dat};

\node[draw, ellipse, minimum width=0.1cm, minimum height=0.25cm, inner sep=0] (E) at (axis cs: 1.2, 0.59) {}; 
\draw (true.south) -- (E); 

\addplot [draw=kit-purple, densely dashed, line width=1] table [
    x=x, y=mean,
] {figures/data/pareto_false_true_4_fir.dat};

\addplot [draw=kit-cyan, densely dashed, line width=1] table [
    x=x, y=mean,
] {figures/data/pareto_true_false_4_fir.dat};

\addplot [draw=kit-purple, line width=1] table [
    x=x, y=mean,
] {figures/data/pareto_false_true_4.dat};

\addplot [draw=kit-cyan, line width=1] table [
    x=x, y=mean,
] {figures/data/pareto_true_false_4.dat};

\end{axis}

\begin{axis}[
width=3.3cm,
height=4cm,
at={(3.4cm,0)},
scale only axis,
ytick distance=1,
minor y tick num=4,
ytick={0, 0.5, 1, 1.5},
yticklabels={},
ymin=0.1,
ymax=1.1,
xmax=1e5,
xlabel=Complexity (rvmb),
xmode=log,
xmin=2e-1,
xminorticks=true,
minor x tick num=8,
xtick={0.1, 1, 10, 1e2, 1e3, 1e4, 1e5},
xticklabels={,$10^0$,,$10^2$,,$10^4$,},
xlabel style={at={(0.50, 0.05)}, anchor=near ticklabel},
grid=major,
minor grid style={thin, gray!20},
no marks,
legend style={
    at={(1,0.05)},
    anchor=south east
},
]

\node[anchor=south west, fill=white, inner sep=2pt] at (axis cs: 0.3, 0.95) {\footnotesize b) 8PAM};

\node[font=\scriptsize, fill=white, inner sep=2pt,align=center] (true) at (axis cs: 500, 0.9) {Joint\\shaping};

\draw [kit-gray, line width=0.7, font=\scriptsize, densely dotted, anchor=south east] (axis cs:2e-1, 0.333) -- (axis cs:1e5, 0.333)
    node[pos=0.75, above, fill=white, inner sep=2pt] {$R_\text{max}^\text{2PAM}=0.\bar{3}$};
\draw [kit-gray, line width=0.7, font=\scriptsize, densely dotted, anchor=south west] (axis cs:2e-1, 0.666) -- (axis cs:1e5, 0.666)
    node[pos=0.25, above, fill=white, inner sep=2pt] {$R_\text{max}^\text{4PAM}=0.\bar{6}$};

\addplot [draw=black!30, line width=1] table [
    x=x, y=mean,
] {figures/data/pareto_true_true_8.dat};

\node[draw, ellipse, minimum width=0.1cm, minimum height=0.25cm, inner sep=0] (E) at (axis cs: 2e4, 0.745) {}; 
\draw (true.east) -- (E); 

\addplot [draw=kit-purple, densely dashed, line width=1] table [
    x=x, y=mean,
] {figures/data/pareto_false_true_8_fir.dat};

\addplot [draw=kit-cyan, densely dashed, line width=1] table [
    x=x, y=mean,
] {figures/data/pareto_true_false_8_fir.dat};

\addplot [draw=kit-purple, line width=1] table [
    x=x, y=mean,
] {figures/data/pareto_false_true_8.dat};

\addplot [draw=kit-cyan, line width=1] table [
    x=x, y=mean,
] {figures/data/pareto_true_false_8.dat};

\end{axis}

\end{tikzpicture}%
    \caption{Pareto fronts of NGMI$\times$rvmb for 4,8PAM with probabilistic shaping and with geometric shaping for FIR filters and for $n$-layer \acp{CNN}.}
    \label{fig:paretos_vs}
\end{figure}
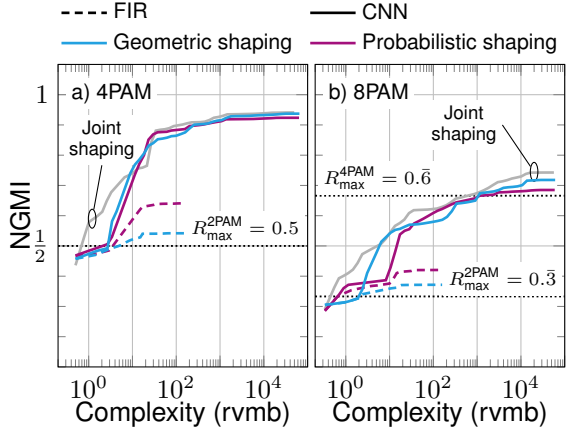

In Fig.~\ref{fig:paretos_vs} we compare the Pareto fronts of geometric and probabilistic shaping. For an FIR-based DSP, probabilistic shaping provides greater gains than geometric shaping. However, once we allow for non-linear DSP, we observe that there is no clear advantage of one over the other. For 4PAM, either geometric or probabilistic shaping is sufficient to achieve a similar performance than joint shaping. In this case, using only geometric shaping can be advantageous, since in this case no \ac{DM} is necessary. Using a \ac{DM} has been proven challenging from the point of view of constellation design and decoding complexity in order to achieve reliable communications~\cite{pas}. For 8PAM, higher rates are obtained by joint shaping. 

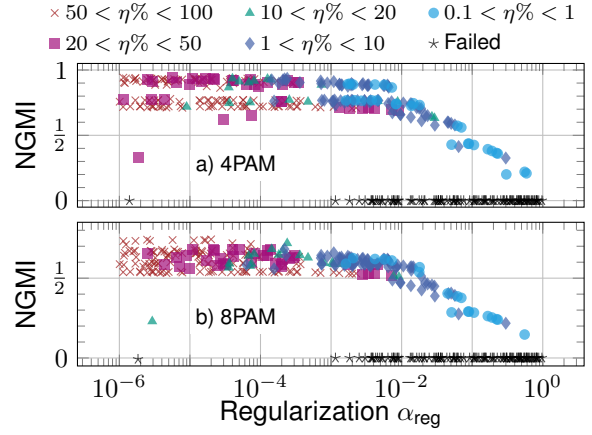
\begin{figure}
    \centering
    \begin{tikzpicture}

\begin{axis}[
width=6.7cm,
height=1.9cm,
at={(0,2.1cm)},
scale only axis,
ytick distance=1,
minor y tick num=4,
ylabel=NGMI,
ylabel style={at={(0.12, 0.5)}, anchor=near ticklabel},
ytick={0, 0.5, 1, 1.5},
yticklabels={$0$, $\frac{1}{2}$, $1$},
ymin=-0.05,
ymax=1.05,
xmode=log,
xminorticks=true,
minor x tick num=8,
xtick={1e-7, 1e-6, 1e-5, 1e-4, 1e-3, 1e-2, 1e-1, 1, 10},
xticklabels={},
xlabel style={at={(0.50, 0.05)}, anchor=near ticklabel},
grid=major,
minor grid style={thin, gray!20},
legend style={
    at={(1, 1.01)},
    anchor=south east,
    legend columns=2,
    legend cell align={left},
    inner sep=1pt,
    draw=none,
    /tikz/every even column/.append style={column sep=0.35cm},
    font=\footnotesize,
    transpose legend,
}
]

\node[anchor=south west, fill=white, inner sep=2pt] at (axis cs: 1e-5, 0.15) {\footnotesize a) 4PAM};

\addplot[
    kit-red,
    only marks,
    mark size=2pt,
    mark=x,
    restrict expr to domain={\thisrowno{2}}{0.5:1},
    restrict expr to domain={\thisrowno{1}}{0.1:1},
    opacity=0.7,
] table {figures/data/reg_lambda_4_rel_cost_15_inf.dat};
\addlegendentry{$50<\eta\%<100$}

\addplot[
    kit-purple,
    only marks,
    mark size=2pt,
    mark=square*,
    restrict expr to domain={\thisrowno{1}}{0:1},
    restrict expr to domain={\thisrowno{2}}{0.2:0.5},
    opacity=0.7,
    draw opacity=0,
] table {figures/data/reg_lambda_4_rel_cost_15_inf.dat};
\addlegendentry{$20<\eta\%<50$}

\addplot[
    kit-green,
    only marks,
    mark size=2.2pt,
    mark=triangle*,
    restrict expr to domain={\thisrowno{2}}{0.1:0.2},
    opacity=0.7,
    draw opacity=0,
] table {figures/data/reg_lambda_4_rel_cost_15_inf.dat};
\addlegendentry{$10<\eta\%<20$}

\addplot[
    kit-blue,
    only marks,
    mark size=2.5pt,
    mark=diamond*,
    restrict expr to domain={\thisrowno{2}}{0.01:0.1},
    opacity=0.7,
    draw opacity=0,
] table {figures/data/reg_lambda_4_rel_cost_15_inf.dat};
\addlegendentry{$1<\eta\%<10$}

\addplot[
    kit-cyan,
    only marks,
    mark size=2pt,
    mark=*,
    restrict expr to domain={\thisrowno{2}}{0.001:0.01},
    opacity=0.7,
    draw opacity=0,
] table {figures/data/reg_lambda_4_rel_cost_15_inf.dat};
\addlegendentry{$0.1<\eta\%<1$}

\addplot[
    black,
    only marks,
    mark size=2pt,
    mark=star,
    restrict expr to domain={\thisrowno{1}}{-1:0.1},
    opacity=0.7,
] table {figures/data/reg_lambda_4_rel_cost_15_inf.dat};
\addlegendentry{Failed}

\end{axis}

\begin{axis}[
width=6.7cm,
height=1.9cm,
at={(0,0)},
scale only axis,
ytick distance=1,
minor y tick num=4,
ylabel=NGMI,
ylabel style={at={(0.12, 0.5)}, anchor=near ticklabel},
ymin=-0.05,
ymax=0.85,
ytick={0, 0.5, 1, 1.5},
yticklabels={$0$,$\frac{1}{2}$,$1$,},
xlabel=Regularization $\alpha_\text{reg}$,
xmode=log,
xminorticks=true,
minor x tick num=8,
xtick={1e-7, 1e-6, 1e-5, 1e-4, 1e-3, 1e-2, 1e-1, 1, 10},
xticklabels={,$10^{-6}$,,$10^{-4}$,,$10^{-2}$,,$10^0$,},
xlabel style={at={(0.50, 0.10)}, anchor=near ticklabel},
grid=major,
minor grid style={thin, gray!20},
]

\node[anchor=south west, fill=white, inner sep=2pt] at (axis cs: 1e-5, 0.15) {\footnotesize b) 8PAM};

\addplot[
    kit-red,
    only marks,
    mark size=2pt,
    mark=x,
    restrict expr to domain={\thisrowno{2}}{0.5:1},
    opacity=0.7,
] table {figures/data/reg_lambda_8_rel_cost_15_inf.dat};

\addplot[
    kit-purple,
    only marks,
    mark size=2pt,
    mark=square*,
    restrict expr to domain={\thisrowno{1}}{0:1},
    restrict expr to domain={\thisrowno{2}}{0.2:0.5},
    opacity=0.7,
    draw opacity=0,
] table {figures/data/reg_lambda_8_rel_cost_15_inf.dat};

\addplot[
    kit-green,
    only marks,
    mark size=2.2pt,
    mark=triangle*,
    restrict expr to domain={\thisrowno{2}}{0.1:0.2},
    opacity=0.7,
    draw opacity=0,
] table {figures/data/reg_lambda_8_rel_cost_15_inf.dat};

\addplot[
    kit-blue,
    only marks,
    mark size=2.5pt,
    mark=diamond*,
    restrict expr to domain={\thisrowno{2}}{0.01:0.1},
    opacity=0.7,
    draw opacity=0,
] table {figures/data/reg_lambda_8_rel_cost_15_inf.dat};

\addplot[
    kit-cyan,
    only marks,
    mark size=2pt,
    mark=*,
    restrict expr to domain={\thisrowno{2}}{0.001:0.01},
    opacity=0.7,
    draw opacity=0,
] table {figures/data/reg_lambda_8_rel_cost_15_inf.dat};

\addplot[
    black,
    only marks,
    mark size=2pt,
    mark=star,
    restrict expr to domain={\thisrowno{1}}{-1:0.1},
    opacity=0.7,
] table {figures/data/reg_lambda_8_rel_cost_15_inf.dat};

\end{axis}

\end{tikzpicture}%
    \caption{Impact of the regularization parameter $\alpha_\text{reg}$ on the \ac{NGMI} and the compression ratio $\eta$, including models that failed to converge.}
    \label{fig:reg_lambda}
\end{figure}

In order to visualize the effect of the regularization coefficient $\alpha_\text{reg}$ on the performance of the detector, we evaluated the \ac{CNN} ensamble with both $\alpha_\text{reg}>0$ and $\alpha_\text{reg}=0$, and we consider the compression factor given by the ratio of the number of \acp{rvmb} for the sparse model and for the dense model defined as $\eta=\frac{\text{rvmb}_{\lambda>0}}{\text{rvmb}_{\lambda=0}}$. In Fig.~\ref{fig:reg_lambda}, we see that $\eta<1\%$ is achievable for 4PAM while still maintaining a close to the maximum \ac{NGMI} for this SNR. For 8PAM, the compression of the models comes with a greater \ac{NGMI} penalty and hence the higher complexity required for good performance as observed previously in Fig.~\ref{fig:paretos_ngmi}. 

\begin{figure}
    \centering
    \input{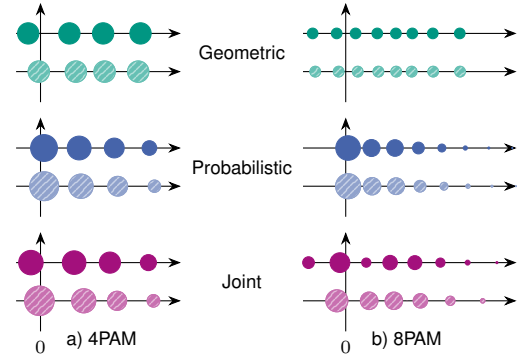}%
    \caption{Trained constellations for low-(hatched) and high-complexity (solid) detectors. The size of the constellation points is proportional to their probability $p_i$.}
    \label{fig:scatter}
\end{figure}

In Fig.~\ref{fig:scatter} we plot some of the trained constellations for low-complexity decoders, with $\text{rvmb}\approx100$ for 4PAM and $\approx1000$ for 8PAM, and for very high complexity $>30000$ \ac{rvmb} for both 4,8PAM. We observe that, for high-complexity detectors, the constellations spread further into the $x<0$ region of the line, following the intuition that a complex enough detector would enable the use of bipolar signals and therefore a more efficient allocation of the transmitted power.

\section{Conclusions}

Through extensive simulation results, we demonstrated that high achievable rates with \ac{DD} are still possible with low-complexity non-linear \ac{DSP} and constellation shaping. Furthermore, our findings indicate that while joint optimization is critical for 8PAM, geometric shaping alone offers a low-complexity alternative for 4PAM, since we avoid the implementation challenges of \acp{DM}. These insights provide a clear framework for optimizing \ac{DD} systems considering the specific cost and power constraints of next-generation optical systems.

\clearpage

\section{Acknowledgements}
This work has received funding from the European Research Council (ERC) under the European Union’s Horizon 2020 research and innovation program (grant agreement No. 101001899).

\printbibliography

\vspace{-4mm}

\end{document}